\documentclass[conference]{IEEEtran}
\IEEEoverridecommandlockouts
\usepackage{cite}
\usepackage{amsmath,amssymb,amsfonts}
\usepackage{eqparbox}
\usepackage{arydshln}
\usepackage{algorithmic}
\usepackage{graphicx}
\usepackage{textcomp}
\usepackage{xcolor}
\usepackage{multirow}
\usepackage{tablefootnote}

\def\BibTeX{{\rm B\kern-.05em{\sc i\kern-.025em b}\kern-.08em
    T\kern-.1667em\lower.7ex\hbox{E}\kern-.125emX}}

\newcommand{\linebreakand}{%
  \end{@IEEEauthorhalign}
  \hfill\mbox{}\par
  \mbox{}\hfill\begin{@IEEEauthorhalign}
}   
    
\begin{document}

\title{Digital Twin-Empowered Smart Attack Detection System for 6G Edge of Things Networks}

\author{
\IEEEauthorblockN{
Yagmur Yigit \IEEEauthorrefmark{1}\IEEEauthorrefmark{2},
Christos Chrysoulas \IEEEauthorrefmark{1}, 
Gokhan Yurdakul \IEEEauthorrefmark{3}, 
Leandros Maglaras \IEEEauthorrefmark{1}, and 
Berk Canberk \IEEEauthorrefmark{1}}
\IEEEauthorblockA{
\IEEEauthorrefmark{1} School of Computing, Engineering and The Build Environment, Edinburgh Napier University, United Kingdom \\
\IEEEauthorrefmark{2} Department of Computer Engineering, Istanbul Technical University, Turkey \\
 \IEEEauthorrefmark{3} BTS Group, Turkey
 \\
Email: \{Y.Yigit, C.Chrysoulas, L.Maglaras, B.Canberk\}@napier.ac.uk, \\ yigity20@itu.edu.tr, yurdakulg@btsgrup.com}
}

\maketitle

\begin{abstract}
As global Internet of Things (IoT) devices connectivity surges, a significant portion gravitates towards the Edge of Things (EoT) network. This shift prompts businesses to deploy infrastructure closer to end-users, enhancing accessibility. However, the growing EoT network expands the attack surface, necessitating robust and proactive security measures. Traditional solutions fall short against dynamic EoT threats, highlighting the need for proactive and intelligent systems. We introduce a digital twin-empowered smart attack detection system for 6G EoT networks. Leveraging digital twin and edge computing, it monitors and simulates physical assets in real time, enhancing security. An online learning module in the proposed system optimizes the network performance. Our system excels in proactive threat detection, ensuring 6G EoT network security. The performance evaluations demonstrate its effectiveness, robustness, and adaptability using real datasets.

\end{abstract}


\section{Introduction}
\label{sec:intro}

The rapid proliferation of the Internet of Things (IoT) is reshaping the global technological landscape. By 2030, the number of IoT devices is predicted to nearly double, reaching over 29.4 billion \cite{statista}. A significant portion of these devices will be interconnected at the edge, forming what is known as the Edge of Things (EoT) network. In this new era, businesses are transitioning to edge deployments, moving closer to end-users and away from traditional data centres. The EoT network encompasses a distributed computing paradigm, where data processing, storage, and analysis occur closer to the data sources, reducing latency and enhancing real-time responsiveness. This meteoric rise is mirrored by the rapid proliferation of devices connected to the edge, significantly increasing the edge network's workload. As more devices and systems connect to the EoT network, the attack surface for hackers expands, providing them with increased opportunities to exploit vulnerabilities and gain unauthorized access to critical systems.

Edge computing is a rapidly growing market, projected to reach USD 3,605.58 billion by 2032 \cite{edgemarket}. The global cost of cybercrime is expected to rise by 69.94 per cent by 2028, reaching an alarming figure of USD 13.82 trillion \cite{costofattack}. These figures underscore the importance of edge-based detection. Cyber threats to EoT networks can lead to data breaches, service disruptions, and even physical harm. Traditional security solutions may be insufficient, emphasizing the need for intelligent and adaptive systems capable of proactive threat detection.

Digital Twin (DT) technology is promising for bolstering security in 6G EoT networks. It enables real-time monitoring and simulation of physical assets, predicting potential security issues or vulnerabilities \cite{twinPort}. Deploying DT processing at the edge allows timely insights into security threats and informed decision-making to bolster network security and ensure optimized performance \cite{ietf}. Additionally, 6G applications demand more edge servers and introduce new attack vectors targeting local infrastructure and users \cite{survey23}. This highlights the need for comprehensive defence strategies in 6G edge networks. Employing multiple detection models can provide a comprehensive solution to address the dynamic nature of network traffic \cite{AutoML}.
To address these challenges, we present a digital twin-empowered smart attack detection system for 6G edge-of-things networks. Leveraging the capabilities of DT and edge computing, our system aims to establish a robust and resilient defence mechanism against cyber threats from IoT devices and edge connection expansion. In our evaluation, we choose the Long Short-Term Memory Autoencoder (LSTM-AE) model for comparison due to its capacity to capture temporal dynamics. We fine-tuned LSTM-AE hyperparameters through rigorous testing to ensure a robust evaluation of the proposed solution.

The key contributions of this article are as follows:
\begin{itemize}
\item We propose a sophisticated smart attack detection system that integrates DT technology into the edge network, enhancing security and enabling proactive threat detection and response for 6G EoT networks.
\item Our system utilizes a dynamic and adaptive approach to update feature selection (FS) and classification methods consistently. This approach ensures optimal performance in identifying and mitigating various 6G EoT network attack types.
\end{itemize}

The paper proceeds with a literature review in Section~\ref{sec:related}, followed by the proposed solution Section~\ref{sec:proposed} and performance evaluation Section~\ref{sec:performance}. We conclude this paper in Section~\ref{sec:conclusion}.

\section{Related Work}
\label{sec:related}

The prominence of IoT and edge technologies has brought about a heightened emphasis on cybersecurity \cite{Chris}. In this section, we review some relevant works that address similar challenges. Mao \emph{et al.} gave a thorough survey of security threats and countermeasures concerning edge computing, caching, and intelligence regarding 6G network edge \cite{survey23}. 
Yao \emph{et al.} explored existing research on intrusion detection systems and proposed innovative detection methods and hybrid system architecture for edge-based industrial-IoT (IIoT) \cite{Network2019}. An anomaly detection framework based on software-defined networking (SDN) is proposed to address the challenge of DDoS attacks on edge devices in distributed and complex environments, utilizing flow information extracted by the edge controller and the GA-XGBoost algorithm for flow classification \cite{Springer22}. Singh \emph{et al.} suggested an edge-based hybrid intrusion detection framework (EHIDF) using machine learning (ML) approaches to detect both known and unknown attacks in the mobile edge computing (MEC) environment \cite{MEC22}. Their EHIDF outperformed previous works with improved accuracy and reduced false alarm rate.

Lee \emph{et al.} presented a lightweight machine learning-based intrusion detection system called IMPACT, designed specifically for resource-constrained IoT devices, utilizing deep auto-encoder and feature abstraction with linear support vector machine (SVM) \cite{Impact20}. Another work introduces a novel privacy-preserving and collusion-resilient identification system called FLACI for EoT, utilizing federated learning to share models instead of raw data among edge nodes \cite{EoT22}. It uses a community detection technique to find collusive groups of attackers and a rating-based mechanism to evaluate the trustworthiness of nodes. Zhang \emph{et al.} addressed the challenge of model poisoning attacks on DT model training and proposed an algorithm called MASTER, which utilizes multi-timescale deep Q-learning networks to optimize the scheduling of local training epochs and devices for accurate forecasting in smart parks \cite{Smartpark23}. This algorithm achieved endogenous security awareness and significantly improved DT model training accuracy and delay in a smart park integrated with DT and 6G edge intelligence. Moreover, the ADRIoT framework,  an innovative anomaly detection framework for IoT networks utilizing edge computing to uncover potential threats swiftly, is presented \cite{ADRIoT}. It employs an edge-assisted architecture, enabling the detection module to run locally on the edge, facilitating prompt detection of IoT-based attacks. A multi-edge collaborative mechanism is designed to pool resources in a local network to address resource limitations.

Although the mentioned studies significantly contribute to cybersecurity in IoT and edge networks, our proposed system presents a distinctive and innovative approach. It creates a dynamic and adaptive security mechanism by integrating DT technology with edge networks. Through real-time analysis and synchronized virtual representations, our system excels in proactive threat detection and mitigation exhibiting a robust and resilient security posture of 6G EoT networks. 

\begin{figure*}[t]
    \centering
    \includegraphics[width=5.5in]{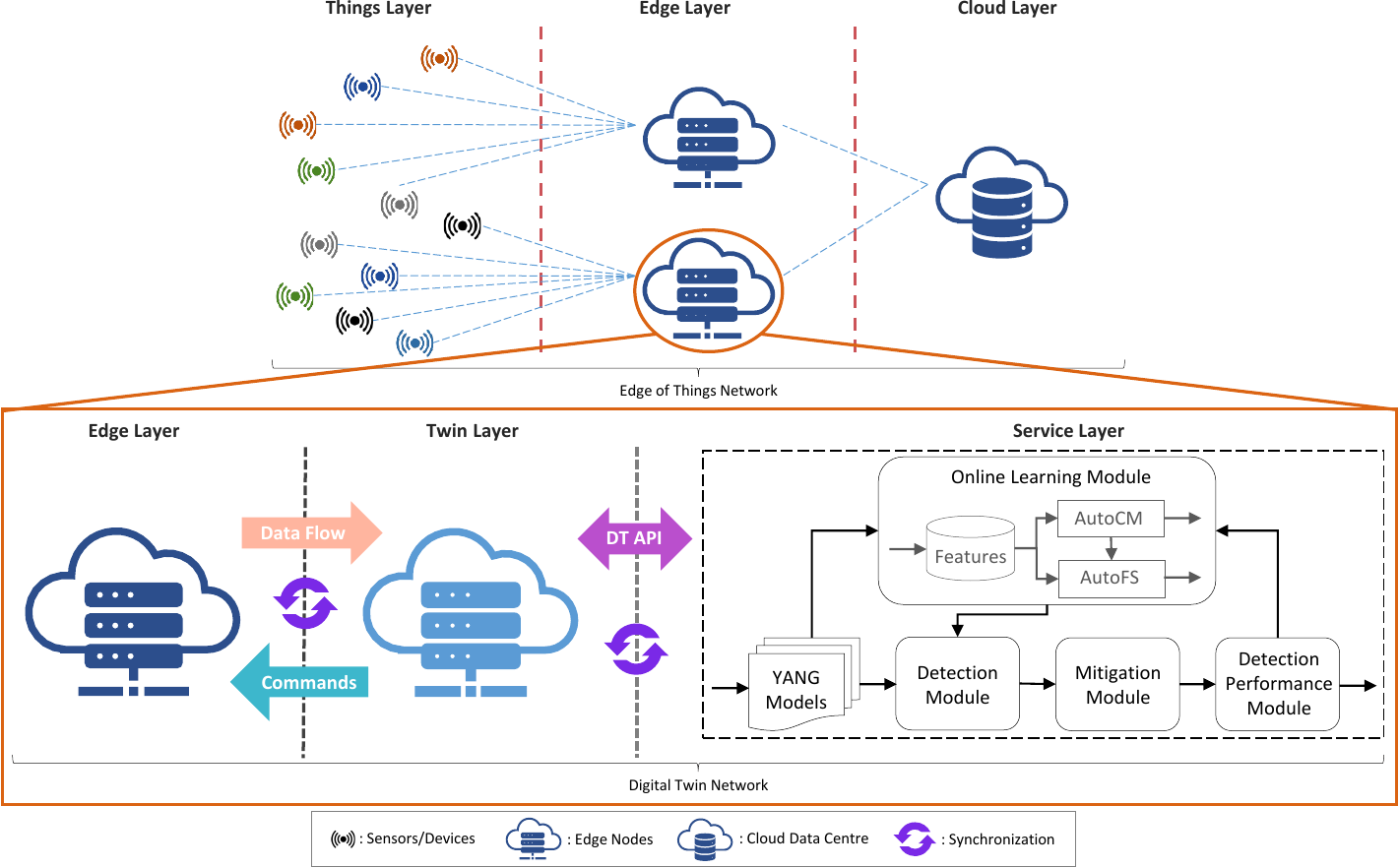}
    \caption{The digital twin-empowered 6G EoT smart attack detection system architecture.}
    \label{fig:system}
\end{figure*}

\section{Proposed 6G EoT System Model}
\label{sec:proposed}

Fig.~\ref{fig:system} depicts the proposed 6G EoT system architecture. It combines two networks. The first network, the EoT network, consists of the things, edge, and cloud layers. 
\textit{The things layer} forms the foundation of the EoT network, encompassing a myriad of interconnected smart devices and sensors that collect data from the physical world. 
\textit{The edge layer} is an intermediary tier between the device and the cloud. It consists of edge computing nodes strategically positioned to the devices they serve. The nodes have relatively higher computational capabilities and perform localized data processing and preliminary analysis. The requirement for constant data transfer to the central cloud is lessened by this layer, which also reduces latency and network congestion. Real-time decision-making, rapid response to emergencies, and low-latency services are all made feasible by edge computing nodes. 
\textit{The cloud layer} represents the traditional centralized cloud infrastructure. Large-scale data centres boast significant computational capabilities and extensive storage capacity in this layer. Additionally, this layer manages resource-intensive operations, complicated data analytics, long-term storage, and other duties that may not be appropriate for the edge layer. The edge layer relieves the strain of delivering all data to the cloud, while the cloud layer assures scalability and thorough analysis, resulting in optimum performance.

The second network is the DT network. In our proposed system, the digital twin of the edge layer is built. We have meticulously constructed a digital twin representation of the edge layer, wherein the entities present within the edge layer mirror the physical elements of our digital twin network. This alignment ensures the edge layer remains closely intertwined with its virtual counterpart. The second layer is \textit{the twin layer}, which is digital replicas of the edge layer entities. This layer enables real-time synchronization and analysis by establishing a smooth connection between the physical and digital worlds. The smart attack detection mechanism is strategically positioned in our third layer of the DT network. As a result, the architecture's overall resilience and dependability are strengthened. This placement allows the system to identify and respond to possible threats and security breaches proactively.

\subsection{Smart Attack Detection}

The functioning of our proposed detection system is delineated through the following sequential steps:
\begin{itemize}
    \item Data generated by the edge node is initially passed through YANG models to facilitate standardized representation and seamless integration with the system.
    \item The data is then transmitted to the detection module, where it undergoes further analysis and evaluation.
    \item Within the detection module, a meticulous assessment is conducted using the system's FS and classification methods to identify potential attacks at the edge node.
    \item In the event of an attack being detected, the mitigation module is promptly activated to neutralize the threat while simultaneously alerting the system administrator regarding the security breach.
    \item In cases where no attack is identified, the detection performance module comes into play. It comprehensively investigates the reliability of the system's classification technique.
    \item The system maintains its current model if the classification method's reliability surpasses a predefined threshold, ensuring continuous operation based on the existing setup.
    \item However, if the classification technique's reliability falls below the predetermined threshold, the detection performance module promptly communicates with the online learning module. This module updates the system's FS and classification methods in near real-time, bolstering its adaptive capabilities and ensuring it remains proficient in identifying and mitigating potential attacks effectively.
\end{itemize}

This approach in the proposed detection system enables proactive threat detection, swift mitigation, and continuous improvement, making it a robust and adaptive solution for safeguarding the edge network against potential security breaches.

\subsubsection{Online Learning Module}

We used our AutoFS \cite{AutoFS} and AutoCM \cite{AutoCM} approaches from our previous works in this module. AutoFS includes five feature selection approaches, while AutoCM contains ten classification algorithms. The general workflow of this module is as follows: After taking a notification from the detection performance module, the online learning module imports one thousand records from the YANG models. Since the obtained data is unlabeled, we employed the labelling method to assign labels to the data and a baseline dataset that contains 65\% of attack samples since attacks are uncommon from our previous work \cite{AutoCM}. 
Unlabeled data undergoes labelling through the application of the labelling algorithm. This process involves augmenting the dataset with one thousand samples from the baseline dataset. Subsequently, ten classification algorithms are employed to train and test their models using two thousand labelled data samples. Finally, the AutoCM selects the most suitable classification method using the final classification method algorithm.

Once the most appropriate classification method is determined, AutoCM transmits this method to AutoFS. AutoFS is responsible for identifying the optimal FS method for the system among five available techniques. The labelled one thousand random data samples are utilized as input for the five FS methods. Each FS method selects the ten most relevant features based on their algorithms. The data, refined by the FS techniques, is employed for training and testing the classification method received from AutoCM.
Subsequently, the performance metrics obtained from the five techniques are forwarded to the final FS algorithm. This algorithm, in turn, determines the best FS method by optimizing the performance metrics for each technique. Once the best FS method is identified, it is the basis for updating the system's FS method and the classification model. This iterative process ensures the system continually adapts to the most effective and efficient FS and classification approach, enhancing its overall performance and accuracy.

\begin{table}[t]
\caption{The Number of Samples Used in Datasets 
\label{tab:dataset}}
\centering
\begin{tabular}{|ll|l|}
\hline
\multicolumn{2}{|c|}{Datasets}                                            & \multicolumn{1}{c|}{Number of Records Used} \\ \hline \hline
\multicolumn{1}{|l|}{\multirow{11}{*}{Edge-IIoT}} & Backdoor Attack       & 1000                                        \\ \cline{2-3} 
\multicolumn{1}{|l|}{}                            & DDoS HTTP Attack      & 500                                         \\ \cline{2-3} 
\multicolumn{1}{|l|}{}                            & DDoS UDP Attack       & 500                                         \\ \cline{2-3} 
\multicolumn{1}{|l|}{}                            & Fingerprinting Attack & 1000                                        \\ \cline{2-3} 
\multicolumn{1}{|l|}{}                            & MITM Attack           & 1000                                        \\ \cline{2-3} 
\multicolumn{1}{|l|}{}                            & Password Attack       & 1000                                        \\ \cline{2-3} 
\multicolumn{1}{|l|}{}                            & Port Scanning Attack  & 1000                                        \\ \cline{2-3} 
\multicolumn{1}{|l|}{}                            & Ransomware Attack     & 1000                                        \\ \cline{2-3} 
\multicolumn{1}{|l|}{}                            & SQL Injection Attack  & 1000                                        \\ \cline{2-3} 
\multicolumn{1}{|l|}{}                            & XSS Attack            & 1000                                        \\ \cline{2-3} 
\multicolumn{1}{|l|}{}                            & Normal                & 18000                                       \\ \hline \hline
\multicolumn{1}{|l|}{\multirow{8}{*}{ToN-IoT}}    & Password Attack       & 1000                                        \\ \cline{2-3} 
\multicolumn{1}{|l|}{}                            & Scanning Attack       & 1000                                        \\ \cline{2-3} 
\multicolumn{1}{|l|}{}                            & XSS Attack            & 1000                                        \\ \cline{2-3} 
\multicolumn{1}{|l|}{}                            & DDoS Attack           & 1000                                        \\ \cline{2-3} 
\multicolumn{1}{|l|}{}                            & Ransomware Attack     & 1000                                        \\ \cline{2-3} 
\multicolumn{1}{|l|}{}                            & Injection Attack      & 1000                                        \\ \cline{2-3} 
\multicolumn{1}{|l|}{}                            & Backdoor Attack       & 1000                                        \\ \cline{2-3} 
\multicolumn{1}{|l|}{}                            & Normal                & 14000                                       \\ \hline
\end{tabular}
\end{table}

The ultimate objective of both the final classification method and FS algorithms is to maximize \emph{\(\sigma_i\)} while simultaneously optimizing \emph{\(\vartheta_i\)} as performance metrics. \emph{\(\sigma_i\)} represents a weighted sum of precision and recall for the \emph{\(i_{th}\)} classification or FS method, whereas \emph{\(\vartheta_i\)} pertains to the detection time associated with the same \emph{\(i_{th}\)} classification or FS method.

\begin{equation}
\begin{split}
       arg\,max \:(\alpha_i \sigma_i + \beta_i \vartheta_i), \:\ i \in [1, 10] \vee [1, 5] 
       \\\\
       \sigma_i = (0.6) \frac{TP}{TP+FN}  +  (0.4) \frac{TP}{TP+FP} , \ \:\: i \in [1, 10] \vee [1, 5] 
       \\\\
       \vartheta_i = t^{end}_{i}\:-\:t^{start}_{i}, \:\: i \in [1, 10] \vee [1, 5] \\
\end{split}
\end{equation}

In Equation 1, \textit{TP} refers to the true positive, \textit{FN} represents the false negative, and \textit{FP} stands for false positive. Additionally, \emph{\(t_{end}\)} denotes the finishing time, while \emph{\(t_{start}\)} represents the starting time.

\subsubsection{Attack Mitigation Module}

Upon successful detection of an attack by the smart attack detection mechanism, this module swiftly comes into action to neutralize the identified threat. This module deploys proactive measures to safeguard the edge nodes and the broader system by leveraging the insights the detection process provides. Through real-time analysis of the attack's characteristics, it formulates targeted countermeasures to mitigate its impact effectively. The malicious traffic is blocked, and the related IP address is added to the suspended IP address list. If the attack is classified as high risk, the system is isolated to the affected edge node. If the attack is classified as mid-high risk, the affected edge node is isolated after taking system admin approval. By integrating such swift and adaptive mitigation strategies, the system can swiftly respond to emerging threats, preserving the integrity and uninterrupted functionality of the 6G edge-of-things network.

\begin{table}[t]
\caption{The Detection Rate Performance
\label{tab:detret}}
\centering
\begin{tabular}{|l|ll|ll|}
\hline
\multicolumn{1}{|c|}{\multirow{2}{*}{Train Dataset}} & \multicolumn{2}{c|}{\multirow{2}{*}{Test Dataset}}                       & \multicolumn{2}{c|}{Detection Rate (\%)}               \\ \cline{4-5} 
\multicolumn{1}{|c|}{}                               & \multicolumn{2}{c|}{}                                                    & \multicolumn{1}{c|}{LSTM-AE} & \multicolumn{1}{c|}{PS} \\ \hline\hline
\multirow{10}{*}{ToN-IoT}                            & \multicolumn{1}{l|}{\multirow{10}{*}{Edge-IIoT}} & Password Attack       & \multicolumn{1}{l|}{91.94}   & 99.46                   \\ \cline{3-5} 
                                                     & \multicolumn{1}{l|}{}                            & Port Scaning Attack   & \multicolumn{1}{l|}{81.79}   & 96.89                   \\ \cline{3-5} 
                                                     & \multicolumn{1}{l|}{}                            & DDoS UDP Attack       & \multicolumn{1}{l|}{92.64}   & 99.24                   \\ \cline{3-5} 
                                                     & \multicolumn{1}{l|}{}                            & XSS Attack            & \multicolumn{1}{l|}{93.27}   & 99.82                   \\ \cline{3-5} 
                                                     & \multicolumn{1}{l|}{}                            & MITM Attack           & \multicolumn{1}{l|}{91.08}   & 99.36                   \\ \cline{3-5} 
                                                     & \multicolumn{1}{l|}{}                            & Backdoor Attack       & \multicolumn{1}{l|}{90.39}   & 99.18                   \\ \cline{3-5} 
                                                     & \multicolumn{1}{l|}{}                            & Fingerprinting Attack & \multicolumn{1}{l|}{87.38}   & 97.02                   \\ \cline{3-5} 
                                                     & \multicolumn{1}{l|}{}                            & SQL Injection Attack  & \multicolumn{1}{l|}{91.86}   & 97.34                   \\ \cline{3-5} 
                                                     & \multicolumn{1}{l|}{}                            & Ransomware Attack     & \multicolumn{1}{l|}{87.26}   & 98.62                   \\ \cline{3-5} 
                                                     & \multicolumn{1}{l|}{}                            & DDoS HTTP Attack      & \multicolumn{1}{l|}{94.15}   & 99.36                   \\ \hline \hline
\multirow{7}{*}{Edge-IIoT}                           & \multicolumn{1}{l|}{\multirow{7}{*}{ToN-IoT}}    & DDoS Attack           & \multicolumn{1}{l|}{90.75}   & 99.75                   \\ \cline{3-5} 
                                                     & \multicolumn{1}{l|}{}                            & Injection Attack      & \multicolumn{1}{l|}{88.92}   & 98.92                   \\ \cline{3-5} 
                                                     & \multicolumn{1}{l|}{}                            & Ransomware Attack     & \multicolumn{1}{l|}{79.82}   & 98.25                   \\ \cline{3-5} 
                                                     & \multicolumn{1}{l|}{}                            & XSS Attack            & \multicolumn{1}{l|}{89.79}   & 98.04                   \\ \cline{3-5} 
                                                     & \multicolumn{1}{l|}{}                            & Backdoor Attack       & \multicolumn{1}{l|}{82.96}   & 97.86                   \\ \cline{3-5} 
                                                     & \multicolumn{1}{l|}{}                            & Scanning Attack       & \multicolumn{1}{l|}{85.24}   & 97.35                   \\ \cline{3-5} 
                                                     & \multicolumn{1}{l|}{}                            & Password Attack       & \multicolumn{1}{l|}{92.58}   & 96.80                   \\ \hline
\end{tabular}
\end{table}

\subsubsection{Detection Performance Module}

This module is vital in assessing the classification method's efficacy within the digital twin-empowered 6G EoT smart attack detection system. Essential metrics like TP and FN are used to evaluate the detection performance. The determination of FN and TP in real-world scenarios where ground truth is often unavailable or challenging to establish is difficult. We address this concern by leveraging our labelling method in the online learning module, which combines labelled and unlabeled data to estimate these values. The following equation is used to measure the reliability of the classification method:

\begin{equation}
\varphi = {1 - \frac{FN}{TP+FN}}
\end{equation}

In Equation 2, \emph{\(\varphi\)} denotes the reliability of the classification method, with a specific focus on the FN metric due to its significance in the data division. In cases where no attack is identified, the detection performance module thoroughly investigates the reliability of the system's classification technique. 
The verification of classification techniques primarily focuses on assessing the system's ability to maintain a low rate of FP. While TP and FN may not change, our system continuously monitors network traffic and evaluates the alerts generated.

\begin{figure*}[t]
    \centering
    \includegraphics[width=5in]{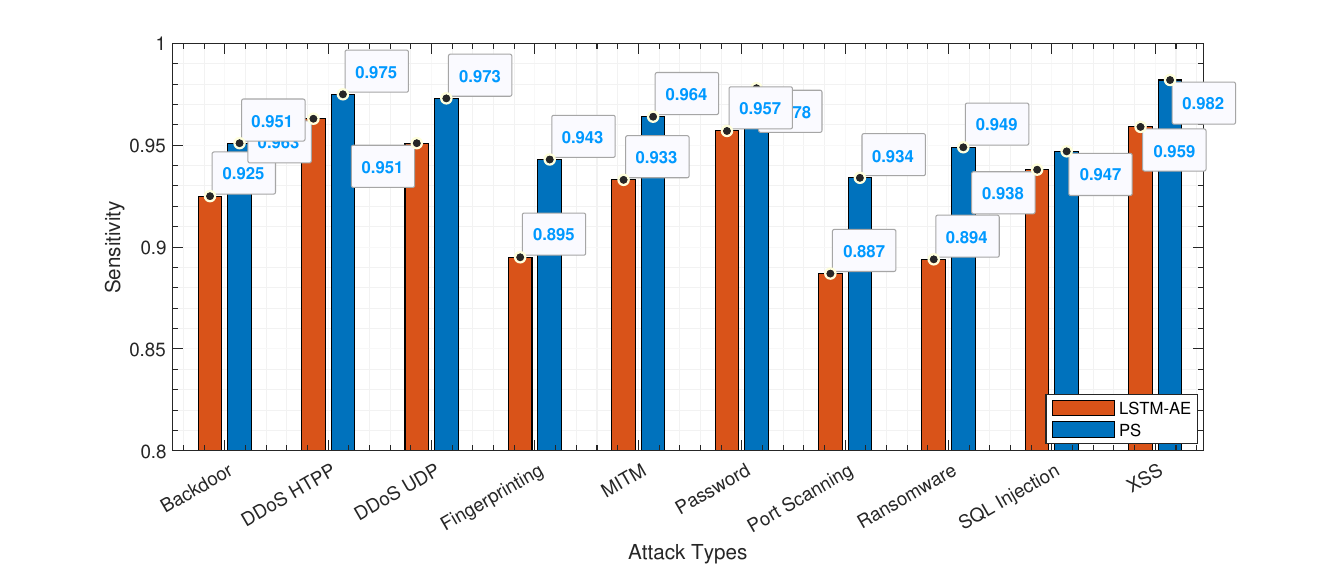}
    \caption{The performance comparison of the Edge-IIoT dataset.}
    \label{fig:EdgeIIoT}
\end{figure*}
\begin{figure*}[t]
    \centering
    \includegraphics[width=5in]{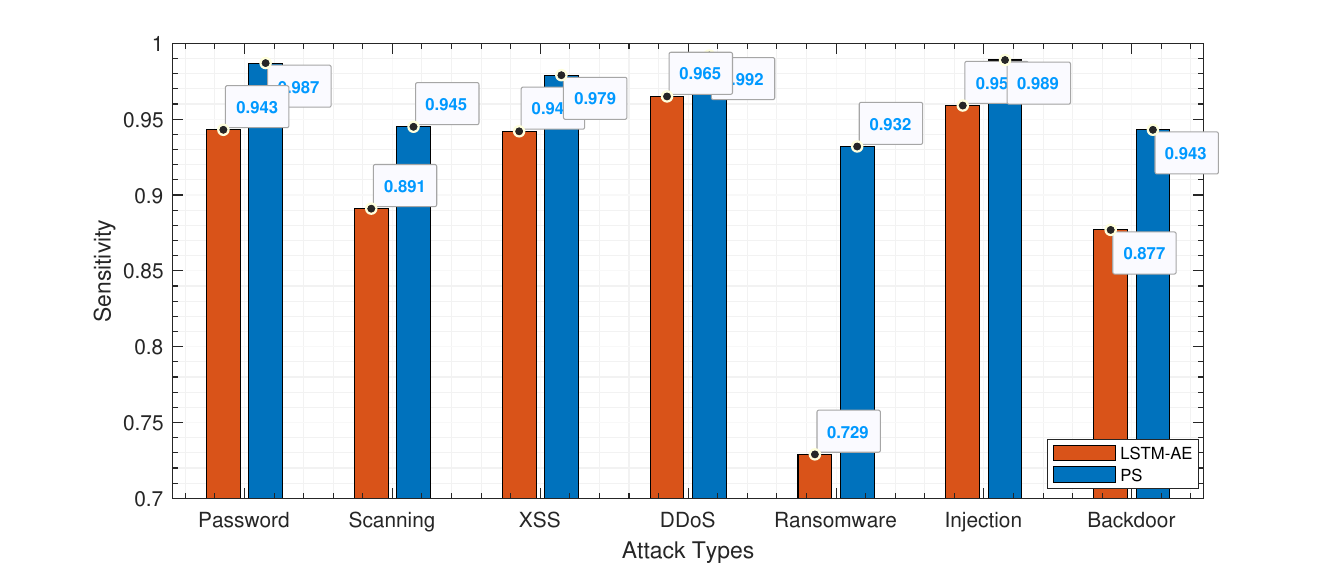}
    \caption{The performance comparison of the ToN-IoT dataset.}
    \label{fig:ToNIoT}
\end{figure*}

The module's decision-making process involves comparing the classification method's reliability against a predefined threshold. 
The reliability threshold for our detection scheme was determined using an adaptive thresholding technique, considering critical factors such as the observed rates of FP and FN over time. When the system shows an elevated FP rate, the threshold is dynamically adjusted to be more stringent, effectively mitigating false alarms. Conversely, if FN rates are a concern, the threshold is appropriately relaxed to enhance detection sensitivity. This approach allows us to maintain an optimal balance between FP and FN, ensuring the reliability and effectiveness of the detection scheme.
The system continues to run using its present model if reliability exceeds the specified threshold, ensuring ongoing and consistent performance based on the current configuration. However, when the classification technique's reliability falls below the predetermined threshold, signalling potential limitations or changes in the system's operational environment, the detection performance module promptly initiates communication with the online learning module. This facilitates near real-time updates to the system's FS and classification methods, empowering the system with adaptive capabilities to identify and mitigate potential attacks effectively. The system improves its overall security and resilience in the constantly changing EoT environment by dynamically altering its defence measures, which keeps it adept in responding to new threats.

\section{Performance Evaluation}
\label{sec:performance}

We built a simple edge network architecture using the NS-3 \cite{ns3}. This network has twelve edge devices in the things layer and two edge nodes in the edge layer. We used the Microsoft Azure DT (ADT) platform to build twin graphs of edge nodes \cite{Azure}. We investigated the performance of our system using Edge-IIoTset \cite{EdgeIIoT}, \cite{d1} and ToN-IoT \cite{ToN_IoT}, \cite{d2} datasets.
The Edge-IIoTset dataset is specifically designed for evaluating IoT and IIoT applications and consists of fourteen attacks targeting connectivity protocols. On the other hand, the ToN IoT dataset was created to assess the effectiveness and efficiency of AI-based cybersecurity applications tailored for next-generation IoTs and industrial IoTs. 
We randomly selected the specific number of samples from these datasets, as seen in Table~\ref{tab:dataset}.
LSTM networks are well-suited for capturing temporal dynamics; therefore, we choose an LSTM network to compare our intrusion detection work.
We conducted a comparison between our proposed solution (PS) and the LSTM-AE utilized in \cite{ADRIoT}. 
To this end, we employed an autoencoder with two encoder layers and two decoder layers. In both the encoder and decoder components, the Dense layer was succeeded by batch normalization and the LeakyReLu activation function. Subsequently, the decoder output features were passed to the LSTM model for further processing. 
The selection of parameters for our LSTM-based model was a result of systematic experimentation and optimization. We conducted tests, cross-validation, and performance evaluations to arrive at the configurations that provided the best trade-off between model complexity and predictive accuracy. We ensured that the LSTM approach is also effective and efficient in comparing our PS.

We employed sensitivity as a performance metric, which represents the ratio of correctly identified attack samples to the total number of samples that should have been identified as attacks.
Initially, we conducted a separate evaluation of the performance results for each dataset. As illustrated in Fig.~\ref{fig:EdgeIIoT} and Fig.~\ref{fig:ToNIoT}, our solution demonstrates superior performance compared to the other approach.
After that, we investigated the detection performance. We trained the initial model with the whole dataset and then tested them with the other dataset. We send the different attacks in order, which is given in Table~\ref{tab:detret}, to test our solution AutoCM and AutoFS performance.

Table~\ref{tab:detret} clearly indicates that our solution exhibits enhanced robustness and adaptability to different attack types. Moreover, it outperforms LSTM-AE regarding attack detection rate, indicating its heightened effectiveness and accuracy in identifying potential threats. 
These achievements underscore our system's heightened effectiveness and accuracy in swiftly identifying and neutralizing potential threats, bolstering the overall security posture of the 6G Edge of Things Networks. Furthermore, the observed superiority of our solution in handling diverse attack scenarios signifies its potential for real-world IoT and IIoT environments, where dynamic security challenges are commonplace. 
These positive outcomes strongly validate the efficacy of our digital twin-empowered smart attack detection system as a proactive and efficient cybersecurity solution, offering a path towards enhanced security and resilience in 6G EoT networks. Furthermore, we assessed the impact of DT on network security by quantifying the reduction in successful attacks and the improvement in incident response times resulting from its implementation. We also scrutinized its resource utilization to ensure it operates efficiently within network constraints while delivering significant security enhancements. These findings underscore the DT's effectiveness as a potent tool for fortifying network security in 6G EoT environments.

\section{Conclusion}
\label{sec:conclusion}
In this paper, we introduced a digital twin-empowered smart attack detection system for 6G Edge of Things networks. Integrating digital twin technology and edge computing enables real-time monitoring and proactive threat detection, bolstering the security of IoT environments. Our system's online learning module ensures continuous improvement by updating feature selection and classification methods, making it adaptable to dynamic attack landscapes. Performance evaluations using real datasets indicate the system's superior performance. The results highlight the system's effectiveness, robustness, and adaptability in detecting diverse attack types, making it a promising solution for securing 6G edge-of-things networks.

\section*{Acknowledgment}
Yagmur Yigit would like to thank the Google DeepMind Scholarship Programme for their support. 

\bibliographystyle{IEEEtran}
\bibliography{main}

\end{document}